# Parametric Modeling of Non-Stationary Signals

Prof Pradip Sircar

Indian Institute of Technology Kanpur

Email address: sircar@iitk.ac.in

**Abstract**

Parametric modeling of non-stationary signals is addressed in this article. We present several models based on the characteristic features of the modelled signal, together with the methods for accurate estimation of model parameters. Non-stationary signals, viz. transient system response, speech phonemes, and electrocardiograph signal are fitted by these feature-based models.

## Introduction

In the model-based signal processing approach, the signal is first fitted into a suitable model, and then this model is utilized for analysis, synthesis and various other signal processing tasks.

The signals that we deal with in real-life situations are often found to be non-stationary in nature [1]. The simplest kind of non-stationary signal is a transient signal which decays with time [2]. Natural speech signals are dominantly non-stationary, and the spectra of such signals often rapidly vary with time [3]. Biological and physiological signals are found to be distinctly non-stationary in their characteristics. The electrocardiograph (ECG) signals show sharp burst-type features [4], and the electroencephalograph (EEG) and electromyography (EMG) signals show rapid variations of amplitude and frequency contents of the signal [5].

Two general linear models for parametric representation of multicomponent non-stationary signals are the time-dependent autoregressive moving-average (ARMA) process [6] and the sum of modulated signals with time-variant amplitude and phase functions [7]. In each of the two representations, the time functions are replaced by some constant-coefficient polynomials of time, and the coefficients are estimated by some suitable techniques. It turns out that the estimation of parameters of each of the two general models is a difficult problem, and the presence of slight noise can even lead to grossly inaccurate results [8].

As an alternative to general models, we propose feature-based models where the model captures the characteristic feature of a non-stationary signal. The complex exponential signal is extensively used for analysis of transient signals [9, 10]. The complex amplitude modulated (AM) and frequency modulated (FM) signal models are employed for analysis and synthesis of speech signals [11, 12, 13]. The AM sinusoidal model with exponential modulation function is fitted into the ECG signal [14, 15]. These are some of the feature-based models for non-stationary signals, which are proposed in recent time.

It is the purpose of the present article to review various parametric models for non-stationary signals. In particular, we consider the models which are proposed based on some characteristic feature(s) of a modelled signal. We mention the salient points of estimation of parameters of each model giving emphasis on the procedures which are common among two or more cases. It is expected that a discussion of various feature-based non-stationary signal models in a unified manner will motivate introduction of more such models in future.

## Complex Exponential Signal

The discrete-time signal $x[n]$ consisting of $M$ complex exponentials is given by

$$x[n] = \sum_{m=1}^{M} A_m \exp(j\varphi_m) \exp(s_m n); \quad n = 0, 1, \cdots, N-1 \quad (1)$$

where $s_m = \alpha_m + j\omega_m$ with $\alpha_m$ negative, and $N$ is the number of signal samples available for processing.

Assuming that the phase $\varphi_m$ are independent and identically distributed (i.i.d.) random variables, and each uniformly distributed over $[0, 2\pi)$, we get the autocorrelation function (acf) $r_x$ of $x[n]$ as

$$r_x[n,k] = E\{x^*[n]x[n+k]\} = \sum_{m=1}^{M} A_m^2 \exp(2\alpha_m n) \exp(s_m k) \quad (2)$$

where $E$ denotes the expectation operation. Since the acf $r_x$ depends on both time $n$ and lag $k$, the modelled signal is a non-stationary signal.

The accumulated acf (aacf) $c_x$ of $x[n]$ is defined as

$$c_x[k] = \sum_{n=n_1}^{n_2} r_x[n,k] = \sum_{m=1}^{M} B_m \exp(s_m k) \quad (3)$$

where $B_m = A_m^2 \sum_{n=n_1}^{n_2} \exp(2\alpha_m n)$, and the time instants $n_1$ and $n_2$ are chosen such that there is no running off the ends of the data block while computing the aacf for the maximum lag.

To estimate the $s_m$-parameters of the modelled signal, we use the computed aacf sequence $\{c_y[k]; k = -J, \cdots, 0, \cdots, J\}$ of the noisy signal samples $y[n] = x[n] + w[n]$, $w[n]$ being noise, in the extended-order Prony's method to obtain a homogenous matrix equation as given below,

$$\mathbf{C}_L \mathbf{a}_L = \mathbf{0} \quad (4)$$

where $\mathbf{C}_L = \left[ c_y[-J + L + i - j]; i = 0, \cdots, 2J - L, j = 0, \cdots, L \right]$,

$\mathbf{a}_L = [1 \, a_1 \, \cdots \, a_L]^T$, and $L$ is much larger than $M$.

We solve the above equation by the singular value decomposition (SVD) technique in the total least squares (TLS) sense to obtain the vector $\mathbf{a}_L$, and form the polynomial equation as follows

$$P(z) = 1 + \sum_{m=1}^{L} a_m z^{-m} = \prod_{m=1}^{L} (1 - z_m z^{-m}) = 0 \quad (5)$$

whose $M$ roots are $z_m = \exp(s_m)$. Once the $s_m$-parameters are estimated, the complex amplitudes $A_m \exp(j\varphi_m)$ are estimated by linear estimation with the noisy signal samples $y[n]$. The $(L-M)$ noise roots of (5), showing corresponding insignificant amplitudes, can be rejected at this stage.

**Complex AM Signal**

The discrete-time signal $x[n]$ consisting of $M$ single-tone AM component signals is represented by

$$x[n] = \sum_{m=1}^{M} A_m \exp(j\varphi_m)\left[1 + \mu_m \exp(j\xi_m n)\right] \exp(j\omega_m n) \quad (6)$$

where the phase $\varphi_m$ are assumed to be i.i.d. random variables, and each uniformly distributed over $[0, 2\pi)$. The acf $r_x$ of $x[n]$ is given by

$$r_x[n,k] = \sum_{m=1}^{M} A_m^2 \left[1 + \mu_m \exp(-j\xi_m n)\right]\left[1 + \mu_m \exp\{j\xi_m(n+k)\}\right] \exp(j\omega_m k) \quad (7)$$

and the aacf $c_x$ of $x[n]$ is expressed as

$$c_x[k] = \sum_{m=1}^{M} B_m \exp(j\omega_m k) + \sum_{m=1}^{M} D_m \exp\left[j(\omega_m + \xi_m)k\right] \quad (8)$$

where $B_m = A_m^2 \left[(n_2 - n_1 + 1) + \mu_m \sum_{n=n_1}^{n_2} \exp(-j\xi_m n)\right]$,

and $D_m = A_m^2 \mu_m \left[\mu_m(n_2 - n_1 + 1) + \sum_{n=n_1}^{n_2} \exp(j\xi_m n)\right]$.

We estimate the circular frequency parameters $\omega_m$ and $\xi_m$ first by spectral analysis utilizing the computed aacf $c_y[k]$ of the noise-corrupted signal samples $y[n]$. Then, the complex amplitudes $A_m \exp(j\varphi_m)$ and modulation indices $\mu_m$ are determined by linear estimation with the sampled sequence $y[n]$.

**Complex FM Signal**

The signal sequence $x[n]$ consisting of $M$ single-tone FM component signals is represented by

$$x[n] = \sum_{m=1}^{M} A_m \exp(j\varphi_m) \exp\left[j\{\omega_m n + \beta_m \sin(\xi_m n)\}\right] \quad (9)$$

where the phase $\varphi_m$ are assumed to be i.i.d. random variables, and each randomly distributed over $[0, 2\pi)$.

The acf $r_x$ of $x[n]$ is expressed as

$$r_x[n,k] = \sum_{m=1}^{M} A_m^2 \exp\left[j2\beta_m \sin(\xi_m k/2)\cos\{\xi_m(2n+k)/2\}\right] \exp(j\omega_m k) \quad (10)$$

which is a function of both time $n$ and lag $k$, and the acf $r_x$ cannot be processed conveniently to estimate the parameters of the model.

We compute the product function $p[k]$ defined and evaluated as shown below:

$$p[k] = \left(x^*[n]x[n+k]\right)_{n=k/2}$$

$$= \sum_{i=1}^{M}\sum_{m=1}^{M} A_i A_m \exp\left[j(\omega_i+\omega_m)k/2\right]\exp\left[j(\varphi_m-\varphi_i)\right]\exp\left[j\{\beta_i\sin(\xi_i k/2)+\beta_m\sin(\xi_m k/2)\}\right]; \quad (11)$$

$$k = 0, \pm 2, \pm 4, \cdots$$

For $M = 2$, the above expression becomes

$$p_2[k] = A_1^2 \sum_{i=-\infty}^{\infty} J_i(2\beta_1)\exp\left[j(2\omega_1+i\xi_1)k/2\right] + A_2^2 \sum_{i=-\infty}^{\infty} J_i(2\beta_2)\exp\left[j(2\omega_2+i\xi_2)k/2\right]$$
$$+ A_{12} \sum_{i=-\infty}^{\infty}\sum_{m=-\infty}^{\infty} J_i(2\beta_1)J_m(2\beta_2)\exp\left[j(\omega_1+\omega_2+i\xi_1+m\xi_2)k/2\right] \quad (12)$$

where $A_{12} = A_1 A_2 \exp\left[j(\varphi_2-\varphi_1)\right] + A_2 A_1 \exp\left[j(\varphi_1-\varphi_2)\right]$, and $J_i(\cdot)$ is the $i$-order Bessel function.

It is apparent that the spectrum of $p[k]$ contains clusters of peaks centred at twice the carrier angular frequencies and at the sum of two carrier angular frequencies. We extract the angular frequency contents and corresponding residues of $p[k]$. Then by residue matching technique, we determine the angular frequencies $\omega_m$ and $\xi_m$, which in turn facilitates the computation of the complex amplitudes $A_m \exp(j\varphi_m)$ and modulation indices $\beta_m$ by solving a linear estimation problem.

**Exponential AM Signal**

The signal sequence $x[n]$ consisting of $M$ single-tone AM component signals with exponential modulations is expressed as

$$x[n] = \sum_{m=1}^{M} A_m \exp(j\varphi_m)\exp\left[b_m\{1-\cos(\xi n - c_m)\}\right]\exp(j\omega_m n) \quad (13)$$

where the phase $\varphi_m$ is i.i.d. random variables, uniformly distributed over $[0, 2\pi)$, and the modulating angular frequency $\xi$ is assumed to be same for all components.

We compute the following two correlation functions $r_{x1}[n]$ and $r_{x2}[n]$ of $x[n]$ defined as

$$r_{x1}[n] = E\{x^*[n]x[n]\} = \sum_{m=1}^{M} A_m^2 \exp\left[2b_m\{1-\cos(\xi n + c_m)\}\right] \quad (14)$$

$$r_{x2}[n] = E\{x^*[-n]x[n]\} = \sum_{m=1}^{M} A_m^2 \exp\left[2b_m\{1-\cos(\xi n + c_m)\}\right]\exp(j2\omega_m n) \quad (15)$$

Assuming that the individual burst features of the modelled signal are separable on the time-axis, we process the two correlation sequences separately.

$$r_{x1m}[n] = A_m^2 \exp\left[2b_m\{1-\cos(\xi n + c_m)\}\right] \quad (16)$$

$$r_{x2m}[n] = A_m^2 \exp\left[2b_m\{1-\cos(\xi n + c_m)\}\right]\exp(j2\omega_m n) \quad (17)$$

It is easy to show that the spectrum of $r_{x1m}[n]$ will have a global maximum at the angular frequency $\omega = 0$, beside local maxima at $\omega = \pm i\xi; i = 1, 2, \cdots$ On the other hand, the spectrum of $r_{x2m}[n]$ will exhibit all the peaks of the spectrum of $r_{x1m}[n]$ with each peak shifted by $\omega = 2\omega_m$. We estimate the circular frequencies $\omega_m$ and $\xi$ by spectral analysis of $r_{x1m}[n]$ and $r_{x2m}[n]$.

To estimate the parameters $A_m$, $b_m$ and $c_m$, we process $r_{x1m}[n]$ as follows:

The correlation function attains its maximum value at

$$n = (\pi - c_m)/\xi \quad (18)$$

This time index is used to estimate $c_m$. Next, the parameters $A_m$ and $b_m$ are estimated by linear estimation with the sequence $\tfrac{1}{2}\ln(r_{x1m}[n])$. The parameter $\varphi_m$ is estimated at last by linear estimation with the sequence $x_m[n]$, which is an individual burst feature assumed to be separable on the time-axis.

**Model Fitting**

Utilizing various feature-based models described above, we now illustrate model fitting of some natural signals which are non-stationary in nature.

**Example 1** Transient Signal

A transient segment of voiced phoneme is fitted by the sum of exponential sinusoidal model which is a pair of complex conjugate exponentials. The original and reconstructed signal segments are plotted in Figure 1. The estimated parameters of the signal are shown in Table 1. The signal reconstruction is done by using the estimated parameters. The example is adopted from [16].

Table 1 Estimated parameters of the complex exponentials

| $\omega$ | $\alpha$ | $A$ | $\varphi$ |
| --- | --- | --- | --- |
| ±3657.6 | −363.97 | 0.5783 | ±5.5229 |
| ±4253.7 | −284.52 | 0.2558 | ±2.4706 |
| ±8234.8 | −235.70 | 0.1834 | ±2.8705 |

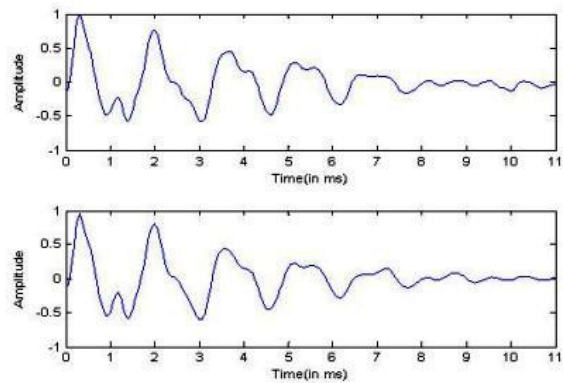

Figure 1 Original (top) and reconstructed (bottom) transient signals

**Example 2** Voiced Speech Phoneme

The sustained vowel sound /ooo/ is fitted by the sum of complex AM sinusoidal signal model. The original and reconstructed speech phonemes are plotted in Figure 2. The estimated parameters of the signal are shown in Table 2. The fitted models appear in complex conjugate pair for the real-valued signal.

Table 2 Estimated parameters of the complex AM signals

| $\omega$ | $\xi$ | $\mu$ | $A$ | $\varphi$ |
|---|---|---|---|---|
| ±5793 | ±1158 | 0.2342 | 1.2393 | ±0.7389 |
| ±1160 | ±3435 | 0.8534 | 0.3609 | ±2.2726 |

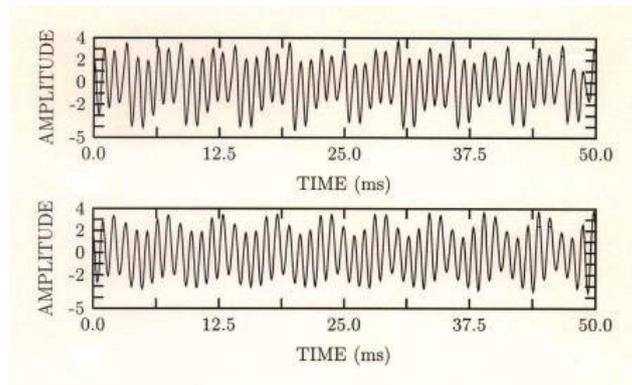

Figure 2 Original (top) and reconstructed (bottom) voiced phoneme /ooo/

**Example 3** Unvoiced Speech Phoneme

The sustained consonant sound /fff/ is fitted by the sum of complex FM sinusoidal signal model. The original and reconstructed speech phonemes are plotted in Figure 3. The estimated parameters of the signal are shown in Table 3. The fitted models appear in complex conjugate pair for the real-valued signal.

Table 3 estimated parameters of the complex FM signals

| $\omega$ | $\xi$ | $\beta$ | $A$ | $\varphi$ |
|---|---|---|---|---|
| ±2886 | ±338 | 0.10 | 0.3371 | ±0.9317 |
| ±5771 | ±162 | 0.15 | 1.0175 | ±1.9398 |
| ±8688 | ±240 | 0.20 | 0.2692 | ±1.0238 |
| ±18738 | ±352 | 0.80 | 0.4265 | ±3.1285 |
| ±1437 | - | - | 0.2872 | ±2.5728 |
| ±4346 | - | - | 0.3083 | ±2.1337 |
| ±7210 | - | - | 0.6955 | ±2.5364 |

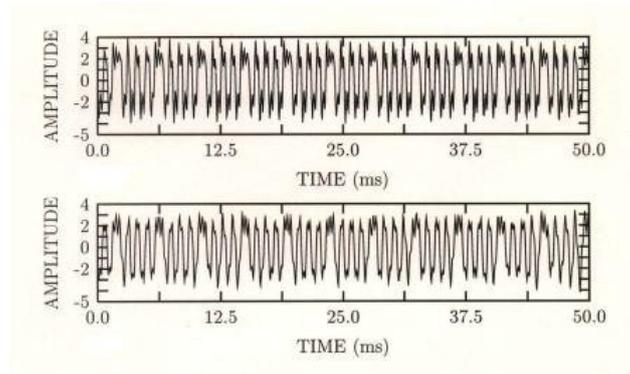

Figure 3 Original (top) and reconstructed (bottom) unvoiced phoneme /fff/

**Example 4** ECG Signal

The ECG signal in lead II configuration is fitted by the sum of exponential AM sinusoidal signal. The P, QRS, T and U waves are separated on the time-axis, and the real signals are converted into the complex analytic signals. The parameters of the modelled signals are estimated, and the estimated parameters are used to reconstruct the real signals. The composite signal is reconstructed by summing the components. The original and the reconstructed ECG signals are plotted in Figure 4, and the estimated parameters are shown in Table 4. The circular frequency $\xi = 2\pi/166$ is same for all components.

Table 4 Estimated parameters of the exponential AM signals fitted in ECG

| Components | $\omega$ | $b$ | $c$ | $A$ | $\varphi$ |
|---|---|---|---|---|---|
| P wave | 3.25583 | 22.6855 | 0.18925 | $3.9499 \times 10^{-18}$ | 0.00 |
| QRS wave | 2.08178 | 44.2553 | 0.30280 | $1.4392 \times 10^{-35}$ | 4.08 |
| T wave | 0.26495 | 2.5724 | 0.07570 | 6.9456 | 1.30 |
| U wave | 4.81711 | 1.8697 | 0.03785 | 3.9012 | 5.20 |

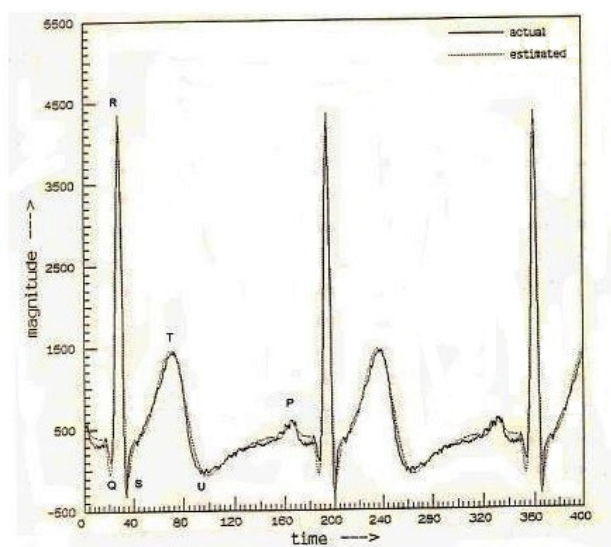

Figure 4 Original (solid) and reconstructed (dotted) ECG signals

## Conclusions

Parametric representations of non-stationary signals by various feature-based models are demonstrated in this article. While parametric representation by a general model essentially leads to an estimation problem which is difficult to solve, parametric representation by a feature-based model often leads to an estimation problem which is tractable in computation. Therefore, the feature-based modeling approach may be the desirable solution for model-based signal processing. However, two important points should be mentioned here. First, when we consider different signals with varied features, we need to look for different models to capture the signal features. Secondly, we need to look for a model which is simple enough to provide a tractable solution of parameter estimation problem. At the same time, the model should not be oversimplified, which may cause a model-mismatch situation.